\documentclass[runningheads]{llncs}
\usepackage{geometry} 
\usepackage[numbers, sort&compress, sectionbib]{natbib} 
\usepackage{amsfonts}  
\usepackage{bm} 

\usepackage{booktabs}       
\usepackage{multirow} 
\usepackage{graphicx}       
\usepackage{array}
\newcolumntype{P}[1]{>{\centering\arraybackslash}p{#1}} 

\usepackage[belowskip=25pt,aboveskip=25pt]{caption} 
\usepackage{subcaption}
\usepackage[font=small]{caption} 
\usepackage[labelfont=bf]{caption} 

\usepackage{etoolbox}
\usepackage{xpatch}
\patchcmd{\thebibliography}{\chapter*}{\section*}{}{} 
\usepackage{hyperref} 

\newcommand{\appendixnumberline}[1]{Appendix\space}
\let\oldappendix\appendix
\makeatletter
\renewcommand{\appendix}{%
  \addtocontents{toc}{\let\protect\numberline\protect\appendixnumberline}%
  \renewcommand{\@seccntformat}[1]{Appendix~\csname the##1\endcsname\quad}%
  \oldappendix
}
\makeatother

\begin{document}

\title{Self-Supervised PPG Representation Learning Shows High Inter-Subject Variability}

\author{Ramin Ghorbani\inst{1}\thanks{Corresponding author~(\email{r.ghorbani@tudelft.nl})}\and
Marcel J.T. Reinders\inst{1}\and
David M.J. Tax\inst{1}}

\institute{Pattern Recognition and Bioinformatics group, Delft University of Technology, Delft, Netherlands}

\maketitle              
\pagestyle{plain}

\begin{abstract}
With the progress of sensor technology in wearables, the collection and analysis of PPG signals are gaining more interest. Using Machine Learning, the cardiac rhythm corresponding to PPG signals can be used to predict different tasks such as activity recognition, sleep stage detection, or more general health status. However, supervised learning is often limited by the amount of available labeled data, which is typically expensive to obtain. To address this problem, we propose a Self-Supervised Learning (SSL) method with a pretext task of signal reconstruction to learn an informative generalized PPG representation. The performance of the proposed SSL framework is compared with two fully supervised baselines. The results show that in a very limited label data setting (10 samples per class or less), using SSL is beneficial, and a simple classifier trained on SSL-learned representations outperforms fully supervised deep neural networks. However, the results reveal that the SSL-learned representations are too focused on encoding the subjects. Unfortunately, there is high inter-subject variability in the SSL-learned representations, which makes working with this data more challenging when labeled data is scarce. The high inter-subject variability suggests that there is still room for improvements in learning representations. In general, the results suggest that SSL may pave the way for the broader use of machine learning models on PPG data in label-scarce regimes.

\end{abstract}
\section{Introduction}
In recent years, wearables such as smartwatches and health trackers, equipped with a photoplethysmography (PPG) sensor, are becoming increasingly popular~\cite{r1}. PPG is a non-invasive, low-cost optical measurement that can measure tissue blood flow over time following each pulse wave ejected from the heart. PPG works on the principle of pulse oximetry, wherein a sensor emits light to the skin and measures the intensity of light that is reflected or transmitted through the skin. Changes in arterial blood volume cause PPG signal variations~\cite{r2,r3}. The cardiac rhythm corresponding to the PPG signal’s periodicity can be used to obtain additional useful information from the users and predict various tasks. Some examples of research on PPG signals are related to Activity Recognition ~\cite{r4}, Heart Rate Estimation~\cite{r5}, Blood Pressure Prediction~\cite{r6}, Biometric Identification~\cite{r7}, Sleep Staging Detection~\cite{r8}, and Atrial Fibrillation Detection~\cite{r9}.

In existing research, analyzing the PPG signals can be broadly categorized into signal processing and machine learning methods. The majority of machine learning solutions for PPG-based tasks utilize fully-supervised learning methods, which can be associated with several limitations. A fully-supervised learning setup usually requires considerable computational resources and time. Additionally, this setup requires large human-annotated datasets for high performance. Typically, obtaining labeled data is very costly and time-intensive, and the amount of labeled data is therefore insufficient in real-world applications, for instance, in the case of heart failure detection or fall detection. When automated detectors have to be trained on this type of problem, a good representation of the data with few numbers of informative features is essential~\cite{r10}. Therefore, it is necessary to address the label-scarcity problem.

One approach to obtain a good informative representation is ‘Self-Supervised Learning’ (SSL). In SSL, two tasks are defined: a ‘pretext’ task and a ‘downstream’ task. The pretext task is the task of learning informative representations by itself. For instance, an auto-encoder tries to precisely reconstruct the input, squeezing the information through a bottleneck layer. It thereby learns a condensed, low-dimensional representation containing all necessary information to reconstruct the input exactly~\cite{r11}. It is assumed that this learned low-dimensional representation reduces the complexity of the data by reducing anomalies and noise and, at the same time, improves the ability to detect patterns in the data simpler and better. Hence, learned representations from the pretext task should be helpful for learning a second-stage classifier on the downstream task, which is the actual task of interest that we want to solve.

The latest research in the field of machine learning shows the potential of SSL for finding generalized and robust representations~\cite{r12,r13,r14,r15}.  The existing works in representation learning are generally concentrated on image-based applications where variations in the data could be visually observed. However, SSL is rarely applied to the field of time series data, especially biosignals. In recent years, some have applied SSL to time series data to show that this method can improve the representation, and they could confirm the potential of self-supervision in capturing important information even in the absence of labeled data. For instance, \citet{r16} introduced an Intra-inter Subject self-supervised Learning (ISL) model customized for ECG signals. Their model integrates medical knowledge into self-supervision to effectively learn from intra-inter subject differences. Their results over different evaluation scenarios showed that the learned representations are information-rich and more generalizable than other state-of-the-art methods for diagnosing cardiac arrhythmias in label-scarce regimes. As another example, \citet{r15} investigated SSL to learn representations from EEG signals. They explored two pretext tasks based on temporal context prediction and contrastive predictive coding on two clinically EEG-relevant downstream tasks. The results show that linear classifiers trained on SSL-learned representations consistently outperform purely supervised deep neural networks in label-scarce regimes while reaching competitive performance when all labels are available. These findings are, however, not yet shown on noisy PPG signals, so it still remains to be shown whether self-supervision can bring improvements over standard supervised approaches on PPG signals.\\

\noindent In this paper, we focus on Human Activity Recognition (HAR) from PPG data. This is gaining interest since PPG data can be easily acquired from any of the widely available wearable devices~\cite{r17}. Researchers have been exploring how SSL techniques can be either extended or explicitly designed for HAR tasks on accelerometer and gyroscope data. However, they have not yet looked into the PPG data specifically. In one of the early pioneering works, \citet{r18} used the task of identifying which signal transformation has been applied to a particular data sample as a pretext task using accelerometer and gyroscope data. The results show that SSL drastically reduces the requirement of labeled activity data, narrowing the gap between supervised and unsupervised techniques for learning meaningful representations. 

\noindent Concluding, to the best of our knowledge, there are currently no studies using SSL on PPG data in label-scarce regimes. Therefore, we present the first detailed analysis of SSL tasks on PPG signals with attention to Activity Recognition as a downstream task. Our main contributions are:

\begin{enumerate}
  \item Proposing a SSL framework for PPG data in label-scarce regimes
  \item Evaluating whether human activity recognition task can be done better when using SSL representations
  \item Investigating the Inter-subject variability in PPG data and exploring how this is captured by the SSL representation
\end{enumerate}
\section{Proposed Framework}
\begin{figure} [ht]
    \centering
    \includegraphics[width=\textwidth]{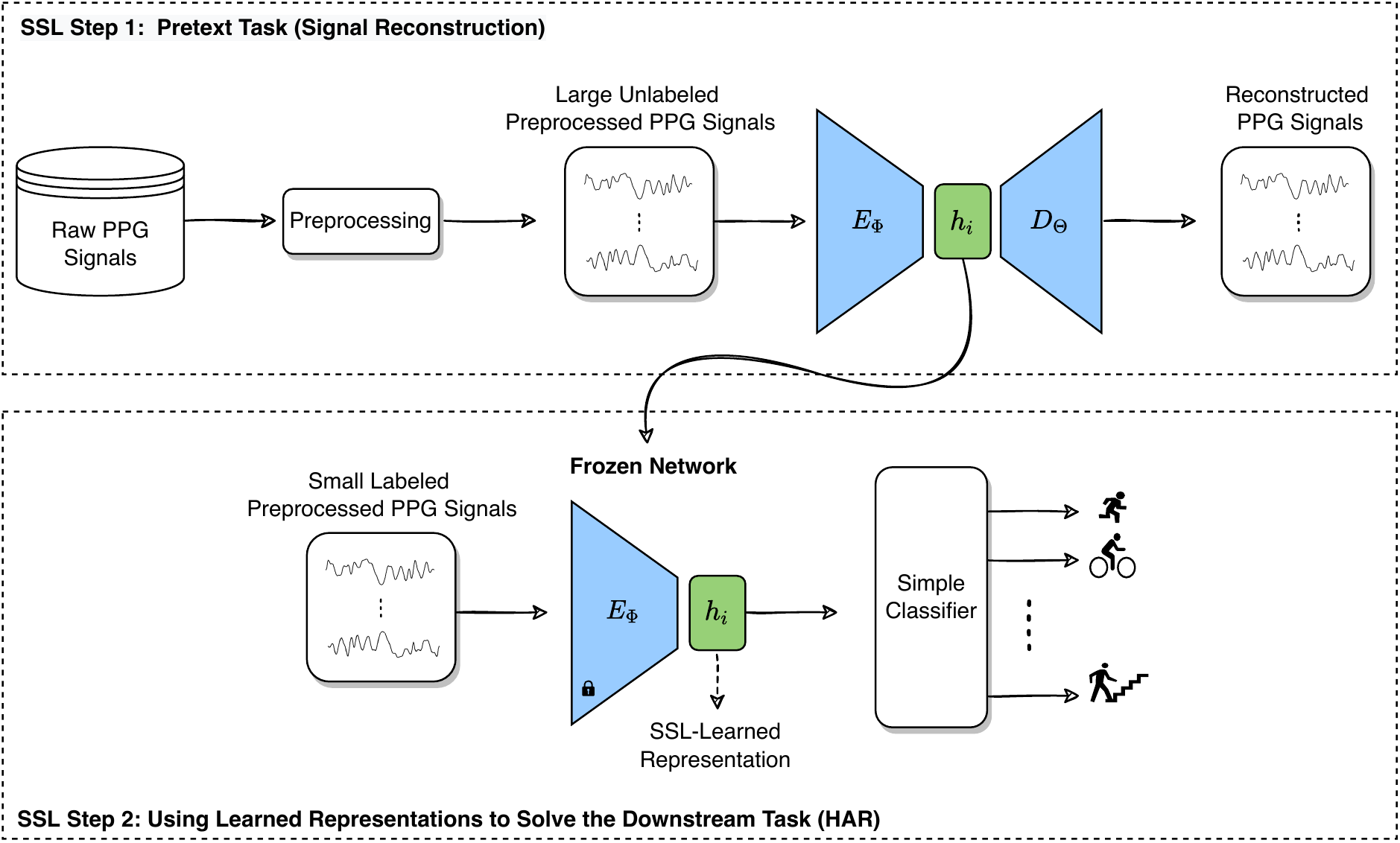}
    \caption{The overall proposed Self-Supervised framework}
    \label{Proposed_Framework}
\end{figure}

An overview of the proposed SSL framework is shown in Figure~\ref{Proposed_Framework}. We use an Autoencoder (AE) to learn a representation of the (unlabeled) data (unsupervised learning). Given an unlabeled dataset $D_U=\left\{ \textbf{x}_{i} \right\}_{i=1}^{N_u}$ where $\textbf{x}_{i}\in \mathbb{R}^{1\times T}$ is a vector of length $T$ and $N_u$ is the number of vectors (samples). The encoder maps each input vector into a latent space representation $\textbf{h}_i=E_\phi(\textbf{x}_i)$ where $\textbf{h}_{i}\in \mathbb{R}^{1\times d}$ where $d< T$. After that, $\textbf{h}_i$ is fed into the decoder component of the model, which follows the same approach to map $\textbf{h}_i$ to the output values $\hat{\textbf{x}}_i=D_\theta (\textbf{h}_i)$ where $\hat{\textbf{x}}_{i}\in \mathbb{R}^{1\times T}$. 
The encoder and decoder are parametrized by $\phi$ and $\theta$, respectively. The AE is trained to minimize the mean squared error between $\textbf{x}_i$ and $\hat{\textbf{x}}_{i}$~\cite{r19}:

\begin{equation}
\textbf{L}_{Total}=1/N_u\sum_{i=1}^{N_u}(1/T\parallel \textbf{x}_{i}-\hat{\textbf{x}}_{i}\parallel^2)
\end{equation}
We have used a combination of Convolutional Neural Network layers (CNN) with the AE architecture, a Convolutional Neural Network AutoEncoder (CNN-AE). The main advantage of using a CNN-AE is a better reconstruction for the PPG signal as it exploits the correlations between time measurements in the PPG signal and thus captures the time-dependent information better.

For the downstream task, we use the original preprocessed PPG as the input to $E_\phi$ with frozen trained weights from the reconstruction task to get the related latent representation $\textbf{h}_{i}$. This $\textbf{h}_{i}$ is used as input to train a simple classifier such as Logistic Regression (LR) or k-Nearest Neighbors (kNN). We do this because when the number of training samples is small, these simple classifiers often outperform more flexible and complex models. The classifier is trained on  $\textbf{h}_{i}$’s from several subjects and tested on a completely new subject.

\section{Experimental Setup}
\subsection{Dataset}
We use the PPG-Dalia public dataset, which was collected by \citet{r20} for the PPG-based heart rate estimation task. This dataset contains recordings of 15 subjects performing daily activities such as \emph{sitting, ascending/descending stairs, playing table soccer, cycling, driving a car, having lunch, walking,} and \emph{working}. The measurements are obtained from wrist and chest-worn devices. Besides the activities, the transient periods between the activities are also recorded. We removed data from subject number 6 due to hardware issues during data recording. Note that \emph{having lunch, driving a car}, and \emph{working} activities are categorized as concurrent or inter-leaved human activities, where actions of multiple activities are carried out simultaneously or where activities contain various activities while their actions can be interleaved in a shuffled manner~\cite{r21,r22}. Therefore, we only considered the remaining five human activities for further study. Detailed information about the dataset is available in Appendix A.

\subsection{Data Preprocessing }
For the pretext task, a band-pass $2^{th} $ order Butterworth filter with low and high frequencies of 0.1 - 6~Hz is applied to the whole PPG signal of each subject individually. The filtered signal is normalized to zero mean and unit variance per subject. The final normalized filtered signals are segmented into a fixed window size of 8 seconds while two successive windows overlap by 6 seconds in the test dataset (this setting is common for PPG data). To increase the training set size, the two successive windows overlap by 7 seconds in the training dataset. For the downstream task, the PPG signals are split into 8 seconds windows while two successive windows overlap by 6 seconds in both training and test datasets. Activity labels are assigned to the corresponding PPG windows based on the available annotations. Detailed information about the labeling process of PPG windows is available in Appendix A.1.

\subsection{Implementation}
\subsubsection{Proposed SSL}
The hyperparameters and the architecture of the proposed CNN-AE are systematically determined by searching through all possible combinations to obtain the best performance. Eventually, we used a CNN-AE architecture deep learning model consisting of three convolution layers, followed by the Exponential Linear Unit (ELU) activation function, Batch Normalization, and MaxPooling layers. The decoder consists of the hidden layers in the reverse order of the encoder section. The Adam optimizer with a learning rate of 0.01, a decay rate of 0.001, and a clip-norm value of 0.9 are used. The batch size is 128, and training runs for 200 epochs. Finally, the parameters of all layers are randomly initialized. To assess the randomness of the deep learning framework, each training process for each test subject is repeated five times. Leave-One-Subject-Out cross-validation (LOSO) is used to evaluate the reconstruction performance. The final implemented CNN-AE model details are available in Appendix B.1.

For the downstream task, two simple classifiers are trained on the SSL-learned representations separately: a Logistic Regression and a kNN classifier (SSL-LR and SSL-kNN, respectively). The SSL-LR is regularized with the L2 penalty term and is solved using LIBLINEAR~\cite{r23}. The SSL-kNN is trained with reweighted neighbors~\cite{r24}, where points are weighted by the inverse of their distance. Therefore, closer neighbors of a query point will have a larger influence than far away neighbors. Due to the different number of training samples which are 2, 5, 10, 50, and 1000 per class, the number of neighbors is selected as 8, 19, 39, 115, and 350, respectively. The LOSO is used to evaluate the AUC performance.

\subsubsection{Comparative Baselines}
The performance of the SSL method is compared with two other baseline models: a simple and a more complex one. The simple baseline model (a typical baseline in SSL research) is trained directly on the original preprocessed PPG representations and consists of the encoder part of the CNN-AE from the pretext task, extended with one classification layer at the end. The encoder part is thus trained on the classification task immediately and not in a self-supervised setting. The Adam optimizer with a learning rate of 0.001 and a clip-norm value of 0.6 are used. The batch size is 128, and training runs for 200 epochs. The more complex baseline is a CNN-LSTM model also trained directly on the original preprocessed PPG representations. The architecture of the complex baseline consists of a convolution layer followed by a hyperbolic tangent function, Batch Normalization, MaxPooling layers, and then a LSTM layer with a hyperbolic tangent activation function, followed by a classification layer at the end. The Adam optimizer with a learning rate of 0.001 and a clip-norm value of 0.6 are used. The batch size is 128, and training runs for 200 epochs. To assess the randomness of these Deep Learning frameworks, each training process for each test subject is repeated five times. Both baseline models use the LOSO to evaluate the AUC performance. The details of the implemented models are available in Appendix B.2 and B.3.

\subsubsection{Biometric Identification (BI) for Exploring Inter-Subject Variability}
If there is a large inter-subject variability, the subjects should be easily discriminated in the representation. To check if a subject can indeed be easily discriminated, we train and evaluate a kNN classifier with reweighted neighbors ($k=20$) per activity. Note that this classifier is not optimized at all on the SSL-learned representations; the kNN fully relies on the metric that is induced by the CNN-AE latent representation $\textbf{h}_{i}$. A good performance of the kNN for the BI task suggests that the learned representation from CNN-AE is heavily biased towards encoding different subjects and not so many other tasks like activities. In this experiment, PPG data is preprocessed with the same steps as the downstream task preprocessing. Afterward, the PPG windows of each activity are selected to sample a separate balanced training set over the subjects. The 4-fold cross-validation (75\% for the training and 25\% for the test set) is used to evaluate the AUC performance.

\section{Results}
\subsection{Pretext Task}
To determine a suitable dimensionality $d$ of the learned representation $\textbf{h}_{i}$, we compute the relative MSE (i.e., MSE in Eq. (1) normalized by the total variance across test subjects’ data) by varying $d$ using the CNN-AE. The results are shown in Table~\ref{tab:table1}. It can be seen that the reconstruction error decreases with increasing dimensionality $d$. As the representation with a lower dimension is more suited for learning with limited labels, we chose $d=64$ when proceeding with the downstream task. Also, later experiments show that  $d=64$ leads to better performance on the downstream task compared to other dimensionalities. Details on the effect of SSL-learned representation dimensionality on the downstream task performance is shown in Appendix C.

\begin{table}[ht]
 \captionsetup{width=0.6\linewidth}
 \caption{Mean Relative MSE results of test subjects (LOSO) for signal reconstruction task by varying $d$ using the CNN-AE }
  \centering
  \begin{tabular}{p{4.5cm}P{4.5cm}}
    \toprule
    Dimension of the $\textbf{h}_{i}$  & Relative MSE Results     \\
    \midrule
    $d=2$       & $0.83\pm 0.02$     \\
    $d=8$       & $0.59\pm 0.03$        \\
    $d=32$      & $0.14\pm 0.03$       \\
    $\bm{d=64}$      & $\bm{0.02\pm 0.00}$       \\
    $d=128$    & $0.00\pm 0.00$       \\
    \bottomrule
  \end{tabular}
  \label{tab:table1}
  \vspace{-4mm}
\end{table}

\subsection{Downstream Task}
\begin{figure}[ht]
    \centering
    \subfloat[\centering]{{\includegraphics[width=7.5cm, height=6cm]{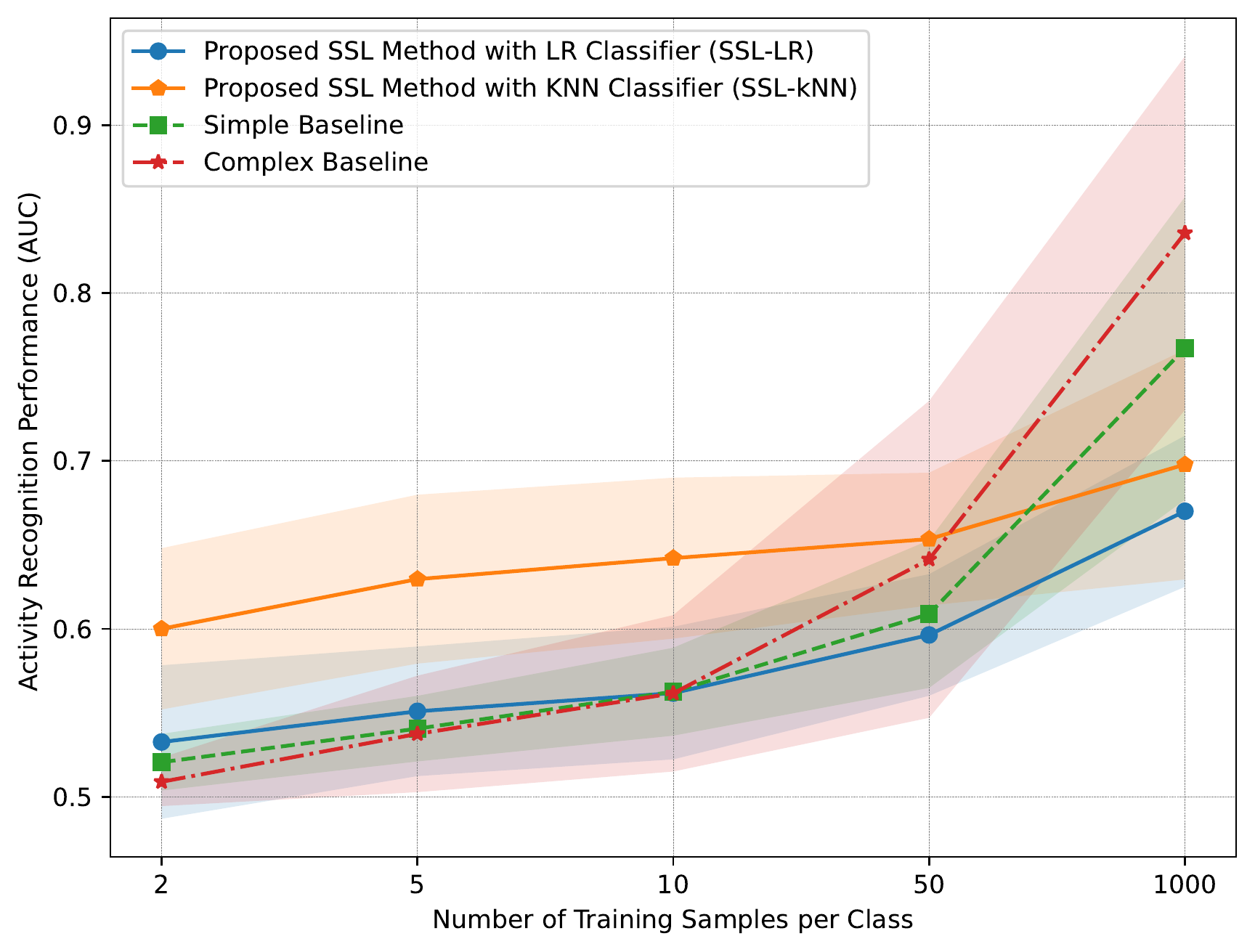} }}%
    \qquad
    \hspace{-5mm}
    \subfloat[\centering]{{\includegraphics[width=7.5cm, height=6cm]{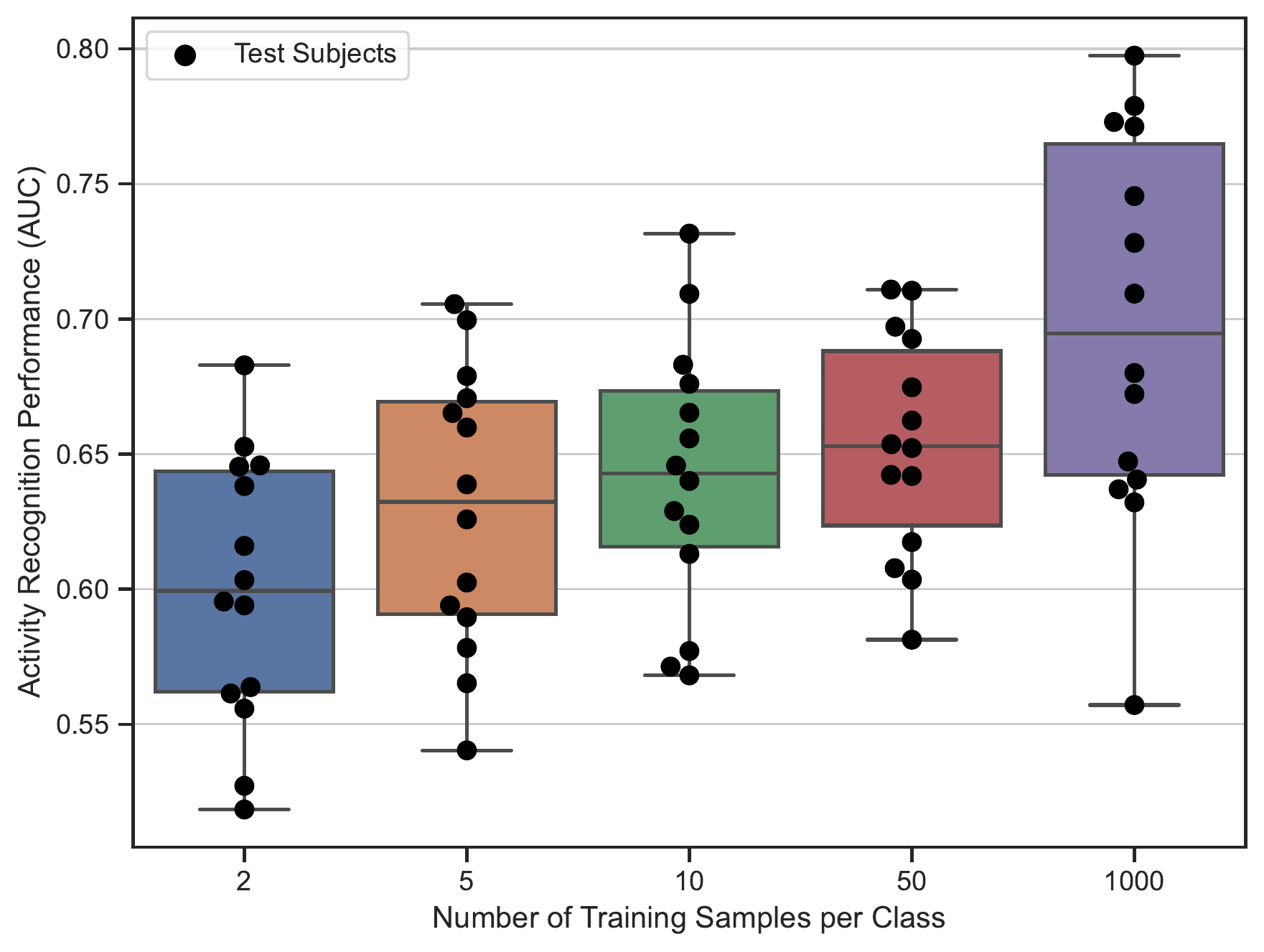} }}%
    \caption{a) Mean AUC performance of test subjects (using LOSO) for Activity Recognition downstream task over the different number of training samples per class. b) The variability among different test subjects in AUC performance for Activity Recognition downstream task using SSL-kNN}%
    \label{fig2}
\end{figure}

In Figure~\ref{fig2}a, we show the AUC performances on the downstream task of predicting activity type for a varying number of training samples per class. The performance of the proposed SSL method is compared with the simple and complex baseline methods. As the number of training samples per class decreases, the performances of all methods drop, confirming the negative influence of when less and less samples with labels are available. The linear SSL-LR model fails to improve the performance compared to the baseline models when a few ($<10$) training samples per class are available. However, SSL-kNN, as a non-linear solution, outperforms the baselines and the SSL-LR in the label-scarce regimes. This suggests that the SSL-learned representation is still too complex for a simple linear solution like LR. One reason for such a behavior could be the high Inter-Subject variability in the SSL-learned representations. Figure~\ref{fig2}b shows the SSL-kNN performance of each of the individual test subjects, for a varying number of training samples per class. It can be observed that the AUC performance can vary between 0.5 and 0.7 for a small training size of $N=2$, and even for very large training sizes of $N=1000$, the AUC still varies between 0.55 and 0.8. This indicates that the data distributions of different subjects vary significantly, thus indicating large inter-subject variability.\\

\noindent To explore the inter-subject variability more deeply, we also investigated the possibility of classifying subjects during each activity using a kNN model. Note that this is now different than our initial domain task; here, we are interested in whether subjects are still separable in the latent representations $\textbf{h}_{i}$ (which, in principle, is undesired when generalizing over subjects). The results of the BI task in Table~\ref{tab:table2} show that subjects can be discriminated perfectly for some activities such as \emph{sitting} or \emph{walking}. It can be seen that there are enough differences among subjects in the original as well as the SSL-learned representations. Moreover, the SSL-learned representation seems to highlight the inter-subject variation as its mean performance is consistently higher than on the original data. This suggests that the AE learned representation is more focused on encoding the different subjects and not so much on the domain tasks of interest, that is, predicting the different activities.
\vspace{-2mm}

\vspace{-2mm}

\begin{table}[h]
    \centering
    \captionsetup{width=0.76\linewidth}
    \caption{Mean AUC Performance of test sets (4-fold cross-validation) for Biometric Identification task }
    \label{tab:table2}
    \begin{tabular}{{p{4.5cm} P{3.5cm} P{3.5cm}}}
    \toprule
    \multirow{2}{*}{Activities} & \multicolumn{2}{c}{\textbf{Input}}                                     \\ \cmidrule{2-3} 
                                & \textbf{Original Representation} & \textbf{SSL-Learned Representation} \\ \midrule
    Sitting  &  $0.84\pm 0.12$   & $0.84\pm 0.01$       \\
    Ascending/Descending Stairs     &  $0.63\pm 0.02$   & $0.64\pm 0.03$       \\
    Playing Table Soccer  &  $0.53\pm 0.01$   & $0.61\pm 0.02$       \\
    Cycling  &  $0.59\pm 0.05$   & $0.61\pm 0.06$       \\
    Walking  &  $0.72\pm 0.02$   & $0.75\pm 0.02$       \\ \bottomrule
    \end{tabular}
    \vspace{-6mm}
\end{table}

\section{Discussion and Conclusion}
We have evaluated the usefulness of self-supervised representation for the activity recognition task when suffering from a label-scarcity in PPG data. The representation is not optimized on the downstream classification task (for which just a few labeled training samples may be available), but it is first optimized to perform a data reconstruction pretext task (for which no supervised information is needed). The results reveal that the SSL method can compete and outperform fully supervised baselines when a kNN model is trained on the SSL-learned representations in label-scarce regimes (with less than 50 samples per class). However, training a simple linear classifier like LR (instead of kNN) is not helpful since the inter-subject variability introduces too much non-linearities in the decision boundaries.  

One should note that in the current study setup, fixed hyperparameters are used across the data regimes for all baseline models. When copious amounts of (unlabeled) data from all subjects would be available, all hyperparameters could be optimized for every different task.

The poor performance of the LR classifier on the SSL-learned representations shows that there is high inter-subject variability. High inter-subject variability makes the generalization more challenging. In this case, a subject-specific model could be a solution for improving the performance over the learned representations. However, training a subject-specific model can be expensive because a large amount of (labeled) data that has to be obtained from each subject. Here, the SSL representation can come to the rescue, as we have shown that this representation can improve performance with respect to the original representation. However, there should be more focus on disentangling the inter- and intra-subject variability. 

This matter opens the door for future research to learn more generalized informative PPG representations while addressing the inter-subject variability problem. For instance, removing the subject-specific factors in order to disentangle the inter-subject variations using a factor disentangling sequential autoencoder~\cite{r25}, or performing contrastive learning among subjects to learn distinctive representations~\cite{r16} can be promising directions in learning informative PPG representations.

\bibliographystyle{unsrtnat}
\renewcommand{\bibname}{\protect\leftline{References}}
\bibliography{ref} 

\newpage
\appendix

\section{Dataset Detailed Information} 
In the PPG-Dalia dataset, each subject followed a defined data collection protocol. The duration of these activities was approximately defined as well. However, since the goal of the data collection was to record data close to the daily-life setting, subjects were instructed to carry out the activities as naturally as possible. An overview of the data collection protocol is given in Table~\ref{tab:tableA}. 

\begin{table}[ht]
\setcounter{table}{0}
\renewcommand{\thetable}{\Alph{table}} 
\centering
\caption{Dalia Datasets Activity Label Information }
\label{tab:tableA}
\begin{tabular}{p{4.2cm}P{2.5cm}p{8cm}} 
\toprule
\multirow{2}{*}{\centering \textbf{Class/Activity}}
  & \textbf{Duration (Min)} & \multicolumn{1}{c}{\multirow{2}{*}{\centering \textbf{Description}}}  \\ \midrule &\\[-0.9em]
Transient Periods           & --             & Before and after each activity, a   transient period was included, in order to arrive at the starting location of   the next activity                                                                                                                                           \\ [3em]
Sitting still               & 10             & Sitting still while reading. (motion artefact-free baseline)                                                                                                                                                                                                                    \\ [1em]
Ascending/Descending stairs & 5              & Climbing six floors up and going down again, repeating this twice                                                                                                                                                                                                               \\ [2em]
Table soccer                & 5              & Playing table soccer, 1 vs. 1 with the supervisor of the data   collection                                                                                                                                                                                                      \\ [2em]
Cycling                     & 8              & Performed outdoors, around research campus, following a defined route of about 2 km length with varying road conditions (gravel, paved)                                                                                                                                         \\ [3em]
Driving car                 & 15             & Started at the parking ground of our research campus and was carried out within the area nearby. Subjects followed a defined route which   took about 15 min to complete. The route included driving on different   streets in a small city as well as driving on country roads \\ [5.5em]
Lunch break                 & 30             & This activity was carried out at the canteen of research campus.  The activity included queuing and fetching food, eating, and talking at the  table                                                                                                                            \\ [3em]
Walking                     & 10             & Walking back from the canteen to the office, with some detour                                                                                                                                                                                                                   \\ [2em]
Working                     & 20             & Subjects returned to their desk and worked on a computer                                                                                                                                                                                                                        \\ [1em] \bottomrule
\end{tabular}
\vspace{-10mm}
\end{table}

\subsection{Labeling Process}
An activity signal containing the label indexes synchronized with the PPG signal is available. After creating the preprocessed PPG windows based on the steps in Section 3, PPG windows should be labeled. At this step, the activity signal is segmented into overlapping windows with the same settings as PPG windows. Later, the activity windows containing two different label indexes are removed along with the corresponding PPG windows. The remaining activity windows have only one activity label index, so this label index is assigned to the corresponding PPG window. 

\section{Deep Learning Models Architecture}
All Deep Learning frameworks are developed and evaluated in python (Version 3.8.8) using Keras API. It should be noted that for the parameters that are not mentioned in the implementation details, the Keras default settings are used. 

\subsection{CNN-AE Model}
The detailed architecture of implemented CNN-AE model for signal reconstruction pretext task is shown in Table~\ref{tab:tableB_1}.

\begin{table}[ht]
\renewcommand{\thetable}{B.1} 
\centering
\caption{The detailed architecture of implemented CNN-AE model }
\label{tab:tableB_1}
\begin{tabular}{p{10cm}|P{2.5cm}|P{1.5cm}}
\toprule
\textbf{Layer (Type)}                               & \textbf{Output Shape} & \textbf{Param \#} \\ \midrule
\multicolumn{3}{l}{\textbf{Encoder}} \\ \midrule
Input Layer                                           & (None, 512, 1)                              & 0                                      \\ 
Conv1D Layer (kernel-size = 32, padding =   “same”)   & (None, 512, 64)                             & 2112                                   \\
Activation Layer                                      & (None, 512, 64)                             & 0                                      \\
Batch Normalization Layer                             & (None, 512, 64)                             & 256                                    \\
Max Pooling1D Layer                                   & (None, 256, 64)                             & 0                                      \\
Conv1D Layer (kernel-size = 32, padding =   “same”)   & (None, 256, 128)                            & 262272                                 \\
Activation Layer                                      & (None, 256, 128)                            & 0                                      \\
Batch Normalization Layer                             & (None, 256, 128)                            & 512                                    \\
Max Pooling1D Layer                                   & (None, 128, 128)                            & 0                                      \\
Conv1D Layer (kernel-size = 32, padding =   “same”)   & (None, 128, 1)                              & 4097                                   \\
Activation Layer                                      & (None, 128, 1)                              & 0                                      \\
Batch Normalization Layer                             & (None, 128, 1)                              & 4                                      \\
Max Pooling1D Layer                                  & (None, 64, 1)                               & 0                                      \\ \midrule
\multicolumn{3}{l}{\textbf{Decoder}} \\ \midrule
Conv1D Layer (kernel-size = 32, padding =   “same”)   & (None, 64, 64)                              & 2112                                   \\
Activation Layer                                      & (None, 64, 64)                              & 0                                      \\
Batch Normalization Layer                             & (None, 64, 64)                              & 256                                    \\
Up Sampling1D Layer                                   & (None, 128, 64)                             & 0                                      \\
Conv1D Layer (kernel-size = 32, padding =   “same”)   & (None, 128, 128)                            & 262272                                 \\
Activation Layer                                      & (None, 128, 128)                            & 0                                      \\
Batch Normalization Layer                             & (None, 128, 128)                            & 512                                    \\
Up Sampling1D Layer                                   & (None, 256, 128)                            & 0                                      \\
Conv1D Layer (kernel-size = 32, padding =   “same”)   & (None, 256, 1)                              & 4097                                   \\
Activation Layer                                      & (None, 256, 1)                              & 0                                      \\
Batch Normalization Layer                             & (None, 256, 1)                              & 4                                      \\
Up Sampling1D Layer (Output Layer)                    & (None, 512, 1)                              & 0                                      \\ \midrule
\multicolumn{3}{l}{\begin{tabular}[c]{@{}l@{}}Total Params: 538,506\\ Trainable Params: 537,734 \& Non-Trainable Params: 772\end{tabular}} \\ \bottomrule
\end{tabular}
\vspace{-0mm}
\end{table}

\subsection{Simple Baseline Model}
The detailed architecture of fully supervised model consists of the encoder part of the CNN-AE model and additional classification layer for HAR downstream task is shown in Table~\ref{tab:tableB_2}. 

\begin{table}[ht]
\renewcommand{\thetable}{B.2} 
\centering
\caption{The detailed architecture of implemented Simple Baseline Model}
\label{tab:tableB_2}
\begin{tabular}{p{10cm}|P{2.5cm}|P{1.5cm}}
\toprule
\textbf{Layer (Type)}                               & \textbf{Output Shape} & \textbf{Param \#} \\ \midrule
Input Layer                                         & (None, 512, 1)        & 0                 \\
Conv1D Layer (kernel-size = 32, padding =   “same”) & (None, 512, 64)       & 2112              \\
Activation Layer                                    & (None, 512, 64)       & 0                 \\
Batch Normalization Layer                           & (None, 512, 64)       & 256               \\
Max Pooling1D Layer                                 & (None, 256, 64)       & 0                 \\
Conv1D Layer (kernel-size = 32, padding =   “same”) & (None, 256, 128)      & 262272            \\
Activation Layer                                    & (None, 256, 128)      & 0                 \\
Batch Normalization Layer                           & (None, 256, 128)      & 512               \\
Max Pooling1D Layer                                 & (None, 128, 128)      & 0                 \\
Conv1D Layer (kernel-size = 32, padding =   “same”) & (None, 128, 1)        & 4097              \\
Activation Layer                                    & (None, 128, 1)        & 0                 \\
Batch Normalization Layer                           & (None, 128, 1)        & 4                 \\
Max Pooling1D Layer                                 & (None, 64, 1)         & 0                 \\
Reshape Layer                                       & (None, 64)            & 0                 \\
Output Layer (activation = “softmax”)               & (None, 5)             & 325               \\ \midrule
\multicolumn{3}{l}{\begin{tabular}[c]{@{}l@{}}Total Params: 269.578\\ Trainable Params: 269,192 \& Non-Trainable Params: 386\end{tabular}} \\ \bottomrule
\end{tabular}
\vspace{-4.5mm}
\end{table}

\subsection{Complex CNN-LSTM Baseline Model}
The detailed architecture of the complex CNN-LSTM fully supervised model for the HAR downstream task is shown in Table~\ref{tab:tableB_3}. 

\begin{table}[ht]
\renewcommand{\thetable}{B.3} 
\centering
\caption{The detailed architecture of implemented CNN-LSTM Baseline Model}
\label{tab:tableB_3}
\begin{tabular}{p{10cm}|P{2.5cm}|P{1.5cm}}
\toprule
\textbf{Layer (Type)}                                                                                              & \textbf{Output Shape} & \textbf{Param \#} \\ \midrule
Input Layer                                                                                                        & (None, 512, 1)        & 0                 \\ [0.5em]
Conv1D Layer (kernel-size = 64, activation   = “tanh”, kernel-regularizer = L2(0.01), bias-regularizer = L2(0.01)) & (None, 449, 32)       & 2080              \\ [1.5em]
Batch Normalization Layer                                                                                          & (None, 449, 32)       & 128               \\
Max Pooling1D Layer (pool-size = 4)                                                                                & (None, 112, 32)       & 0                 \\
Dropout (rate = 0.5)                                                                                               & (None, 112, 32)       & 0                 \\ [0.5em]
LSTM   Layer (kernel-regularizer = L2(0.01), bias-regularizer = L2(0.01))                                          & (None, 32)            & 8320              \\ [1.5em]
Output Layer (activation = “softmax”)                                                                              & (None, 5)             & 165               \\ \midrule 
\multicolumn{3}{l}{\begin{tabular}[c]{@{}l@{}}Total Params: 10,693\\ Trainable Params: 10,629 \& Non-Trainable Params: 64\end{tabular}}                         \\  \bottomrule
\end{tabular}
\end{table}

\section{Effect of SSL-Learned Representation Dimensionality on Downstream Task Performance}
The effect of SSL-learned representation dimensionality on the downstream task performance using SSL-kNN is shown in Table~\ref{tab:tableC}. It can be seen that $d=64$, which has the best mean relative MSE results over test subjects, leads to the best downstream task performance when limited labeled data is available.

\begin{table}[ht]
\renewcommand{\thetable}{C} 
\centering
\captionsetup{width=0.87\linewidth}
\caption{SSL-kNN AUC performance on Activity Recognition downstream task based on different SSL-learned representation dimensionality, showing the effect of using SSL-learned representation with different dimensionality}
\label{tab:tableC}
\begin{tabular}{p{3cm}P{2cm}P{2cm}P{2cm}P{2cm}P{2cm}}
\toprule
\multirow{2}{*}{Samples Per Class} & \multicolumn{5}{c}{Dimensionality of SSL-Learned Representation $\textbf{h}_{i}$}                             \\ \cmidrule{2-6} 
                                   & $d=2$                & $d=8$                & $d=32$                & $d=64$                & $d=128$                \\ \midrule
2                                  & $0.50\pm0.06$ & $0.54\pm0.04$ & $0.57\pm0.05$ & $0.60\pm0.05$ & $0.57\pm0.04$ \\
5                                  & $0.52\pm0.05$ & $0.58\pm0.04$ & $0.59\pm0.06$ & $0.62\pm0.05$ & $0.60\pm0.05$ \\
10                                  & $0.52\pm0.04$ & $0.58\pm0.04$ & $0.61\pm0.06$ & $0.64\pm0.05$ & $0.61\pm0.04$ \\
50                                 & $0.55\pm0.03$ & $0.58\pm0.04$ & $0.64\pm0.04$ & $0.63\pm0.04$ & $0.62\pm0.03$ \\
1000                               & $0.57\pm0.05$ & $0.62\pm0.05$ & $0.67\pm0.08$ & $0.69\pm0.07$ & $0.67\pm0.08$  \\ \bottomrule
\end{tabular}
\end{table}

\end{document}